\documentclass[conference]{IEEEtran}

\usepackage{amsmath}
\usepackage{amsfonts}
\usepackage{graphicx}
\usepackage{cite}
\usepackage{multirow}
\usepackage{algorithm,algorithmic}

\newcommand{\norm}[1]{\left\lVert#1\right\rVert}

\title{A Relax-and-Round Approach to Complex Lattice Basis Reduction}

\author{\IEEEauthorblockN{Marius Arvinte and Ahmed H. Tewfik}
	\IEEEauthorblockA{Department of Electrical and Computer Engineering\\
		University of Texas at Austin\\
		Austin, Texas 78712\\
		Email: arvinte@utexas.edu}}
	
\begin{document}
	
	\maketitle
	
	\begin{abstract}
		We propose a relax-and-round approach combined with a greedy search strategy for performing complex lattice basis reduction. Taking an optimization perspective, we introduce a relaxed version of the  problem that, while still nonconvex, has an easily identifiable family of solutions. We construct a subset of such solutions by performing a greedy search and applying a projection operator (element-wise rounding) to enforce the original constraint. We show that, for lattice basis reduction, such a family of solutions to the relaxed problem is the set of unitary matrices multiplied by a real, positive constant and propose a search strategy based on modifying the complex eigenvalues.
		
		We apply our algorithm to lattice-reduction aided multiple-input multiple-output (MIMO) detection and show a considerable performance gain compared to state of the art algorithms. We perform a complexity analysis to show that the proposed algorithm has polynomial complexity.
	\end{abstract}
	
	\section{Introduction}
	\label{section_intro}
	
	Lattice basis reduction is a key component in the class of lattice-reduction (LR) aided MIMO detection algorithms \cite{lraided_basic}, as well as the recently introduced class of Integer Forcing receivers \cite{if_basic} shown to achieve the maximal diversity order using only linear operations and to close the performance gap between Maximum Likelihood and linear detection.
	
	Lattice basis reduction methods have been extensively studied long before their application to MIMO detection and can be largely categorized as approximate or exact solutions. Minkowski introduced the first approximate algorithm in \cite{minkowski_superold}. In \cite{lattice_book} it was shown that exactly solving the lattice basis reduction problem is equivalent in complexity to solving the closest vector problem, meaning that at least hyper-exponential complexity is required.
	
	The LLL algorithm \cite{lll_regular} is one of the few polynomial complexity algorithms that aims to approximately solve the lattice basis reduction and has been used as the default method in a large part of the LR-aided detection and Integer Forcing literature \cite{lraided_basic, lraided_sic, if_basic}.
	
	Reference \cite{lll_book} presents an extensive survey of the practical performance and running time of the LLL algorithm and reveals that the average case perfomance of the LLL outperforms its theoretical guarantees. Recently, the authors in \cite{boostedlll} have further improved the performance of the LLL algorithm by extending size reductions to length reductions using the parallel nearest plane algorithm. For complex-valued lattices the authors in \cite{clll} introduce the CLLL algorithm that uses the same principles as LLL applied to the complex number field.
	
	Our proposed algorithm relies on treating the lattice basis reduction as a constrained optimization problem. This type of approach was recently used in \cite{slowest_desc}, where the authors relax the integer constraint and use the slowest descent method to construct a set of candidate column vectors for the reduced basis. In constrast, our approach seeks to preserve the overall structure of the reduced basis as orthogonal as possible, instead of constructing it from separate vectors.
	
	Our proposed algorithm operates on complex matrices, but with different core principles than those of the CLLL algorithm. We pose the lattice basis reduction problem as a nonconvex optimization problem with integer constraints and combine a greedy search strategy with a relax-and-round approach inspired by the framework in \cite{boyd_noncvx}. 
	
	The main idea in our approach is to find a suitable factorization of the input channel matrix $\mathbf{H}$ by running a relax-and-round procedure multiple times, with different starting points in $\mathbb{C}^{\textrm{N} \times \textrm{N}}$, exploiting the nonlinear effects of the rounding operation. These effects cause the feasible solutions to be sensitive to the starting point, allowing us to construct a collection of factorizations, out of which a winner is selected.
	
	Our contributions are two-fold: formulating a relaxed version of the lattice basis reduction problem that is solvable exactly and introducing a search heuristic (instead of randomized guesses e.g. as in \cite{boyd_noncvx}) based on the eigenvalues of a complex matrix. To the best of our knowledge, this is the first time such a framework is used for lattice basis reduction.
	
	The algorithm is proposed for use and its performance is simulated in a lattice-reduction aided MIMO detection scenario. The results demonstrate a considerable perfomance gain in the bit error rate (BER) versus signal-to-noise ratio (SNR) domain, while preserving a polynomial time and space complexity. Furthermore, our proposed algorithm is shown to be highly suitable for a parallelized implementation.
	
	The rest of the paper is organized as follows. Section \ref{section_sys} presents the system model and the lattice basis reduction problem. Section \ref{section_prop_alg} describes the proposed algorithm, while Section \ref{section_perf} discusses the choice of numerical parameters, the simulated performance and the complexity. Section \ref{section_conc} concludes the paper.
		
	\textit{Notation}: We use bold uppercase and lowercase characters to denote matrices and column vectors, respectively. Unless otherwise stated, all values are assumed to be complex and all matrices square. The operators $(\cdot)^{\textrm{H}}$ and $(\cdot)^{-1}$ denote the Hermitian and inverse of a square matrix, respectively, $\det$ is the determinant, $( \cdot{} )_{k,k}$ represents the $k$-th diagonal element of a square matrix and $\lceil \cdot \rfloor$ represents the elementwise rounding to nearest integer operator.
	
	\section{System Model}
	\label{section_sys}
	We consider a downlink single-user MIMO scenario, characterized by the narrowband, instantaneous equation: 
	
	\begin{equation}
		\mathbf{y = Hx + n},
	\end{equation}
	
	\noindent where $\mathbf{H} \in \mathbb{C}^{\textrm{N} \times \textrm{N}}$ is the MIMO channel matrix, $\mathbf{x}$ and $\mathbf{y}$ are the transmitted and received signals respectively and $\mathbf{n}$ is the complex, additive, white Gaussian noise. Lattice-reduction aided MIMO detection requires that $\mathbf{x}$ belong to a signal constellation satisfying the lattice properties \cite{lraided_basic}, such as the widespread QAM constellations.
	
	The purpose of lattice reduction in the MIMO detection scenario can be formulated as a constrained optimization problem that consists of finding a factorization of $\mathbf{H}$ in the form \cite{lraided_basic}
	
	\begin{equation}
		\label{basic_factorization}
		\begin{split}
			\mathbf{H} & = \mathbf{QZ}, \\
			& \text{s.t.} \ \mathbf{Z} \in (\mathbb{Z}+j\mathbb{Z})^{\text{N} \times \text{N}},
		\end{split}
	\end{equation}
	
	\noindent where the optimal $\mathbf{Q}$ is the solution to the following optimization problem:
	
	\begin{equation}
		\label{opt_criteria}
		\mathbf{Q} = \underset{\mathbf{Q}_0}{\mathrm{argmin}} \max_k \ (\mathbf{Q}_0^{-1} \mathbf{Q}_0^{-\textrm{H}})_{k,k}.
	\end{equation}
	
	We call the function $c(\mathbf{Q}_0) = \max\limits_k (\mathbf{Q}_0^{-1} \mathbf{Q}_0^{-\textrm{H}})_{k,k}$ the \textit{cost function}.
		
	In a MIMO detector that uses Zero-Forcing, the diagonal entries of the matrix $\mathbf{Q}_0^{-1} \mathbf{Q}_0^{-\textrm{H}}$ represent the per-stream noise amplification factor after inverting the $\mathbf{Q}_0$ matrix at the receiver side. This, coupled with the fact that in such a scenario the overall BER performance is dominated by the stream with lowest post-equalization SNR \cite{lr_determinant_discussion}, justifies the min-max type of optimization in (\ref{opt_criteria}).
	
	An alternative measure for the quality of the reduced basis is the orthogonality defect $od(\mathbf{Q}_0)$ defined as
	
	\begin{equation}
		\label{od_metric}
		od(\mathbf{Q}_0) = \frac{\prod_k \norm{\mathbf{q}_{0k}}^2}{\lvert \det \mathbf{Q}_0 \rvert}.
	\end{equation}
	
	While this metric is not the direct target of our optimization problem, it is still a very good indicator for the quality of the reduced basis \cite{lattice_book} and we investigate it when simulating our algorithm.
	
	Once the optimization problem is solved and given that the signal $\mathbf{x}$ is drawn from a -- potentially shifted and scaled -- finite complex lattice, we can write the equivalent transmitted signal as $\mathbf{s = Zx}$, where $\mathbf{s}$ now belongs to the infinitely extended complex integer lattice. The equivalent MIMO equation is:
	
	\begin{equation}
		\mathbf{y = Qs + n}.
	\end{equation}
	
	We assume that the receiver uses the Zero-Forcing algorithm to recover the signal $\mathbf{s}$, although this can be extended to other well-known heuristics, such as MMSE or search-based algorithms \cite{lraided_sic}, but these are out of scope for this paper. After inverting the $\mathbf{Q}$ matrix, the estimate of $\mathbf{s}$ is obtained as:
	
	\begin{equation}
		\tilde{\mathbf{s}} = \lceil \mathbf{Q}^{-1} \mathbf{y} \rfloor.
	\end{equation}
	
	The original signal $\mathbf{x}$ is then recovered by solving the linear system $\tilde{\mathbf{s}} = \mathbf{Zx}$ either through inverting the matrix $\mathbf{Z}$ or other methods, such as Gaussian elimination.
	
	\section{The Proposed Algorithm}
	\label{section_prop_alg}
	
	Our main idea is to indirectly minimize the cost function $c(\mathbf{Q})$ by first relaxing the integer constraint of the optimization problem and then proposing a modified version that, while still nonconvex, has an achievable global minima for which we can identify a family of feasible solutions. 
	
	Selecting feasible solutions of the relaxed problem is done by a greedy search procedure by using various starting points $\mathbf{Q}_0$, where we parametrize $\mathbf{Q}_0$ by its eigenvalues and a scalar amplitude factor.
	
	Since we require the factorization in (\ref{basic_factorization}) to be exact, we can drop $\mathbf{Q}_0$ and reformulate (\ref{opt_criteria}) as a constrained optimization problem in the variable $\mathbf{Z}_0$ as
	
	\begin{equation}
	\label{first_opt_form}
		\begin{split}
			\mathbf{Z} = \underset{\mathbf{Z}_0}{\mathrm{argmin}} & \max_k (\mathbf{Z}_0 \mathbf{H}^{-1} \mathbf{H}^{-\textrm{H}} \mathbf{Z}_0^\textrm{H})_{k,k}, \\
			& \textrm{s.t.} \ \mathbf{Z}_0 \in (\mathbb{Z}+j\mathbb{Z})^{\textrm{N} \times \textrm{N}}, \\
			& \quad \ \textrm{rank}(\mathbf{Z}_0) = N.
		\end{split}
	\end{equation}
	
	Let $\mathbf{G} = \mathbf{H}^{-1} \mathbf{H}^{-\textrm{H}}$ and notice that $\mathbf{G}$ is a positive definite matrix as long as $\mathbf{H}$ has full rank. Furthermore, noticing that $(\mathbf{Z}_0 \mathbf{G} \mathbf{Z}_0^\textrm{H})_{k,k} = \mathbf{z}_{0,k} \mathbf{G} \mathbf{z}_{0,k}^\textrm{H}$, where $\mathbf{z}_{0,k}$ is the $k$-th line of the $\mathbf{Z}$ matrix we can rewrite (\ref{first_opt_form}) as
	
	\begin{equation}
	\label{second_opt_form}
		\begin{split}
			\mathbf{Z} = \underset{\mathbf{Z}_0}{\mathrm{argmin}} & \max_k \mathbf{z}_{0,k} \mathbf{G} \mathbf{z}_{0,k}^\textrm{H}, \\
			& \textrm{s.t.} \ \mathbf{z}_{0,i} \in (\mathbb{Z}+j\mathbb{Z})^{1 \times \textrm{N}}, \\
			& \quad \ \textrm{rank}(\mathbf{Z}_0) = N,
		\end{split}
	\end{equation}
	
	\noindent where $i$ ranges from 1 to $N$.
	
	The next step consists in a reformulation of the problem in (\ref{second_opt_form}). Note that if we relax the integer constraint in the current form, the problem does not have a solution with zero cost. Instead, we can set each of the $\mathbf{z}_{0,i}$ vectors equal to a real, positive value $\epsilon_i$ times the $i$-th eigenvector of $\mathbf{G}$, leading to the cost function
	
	\begin{equation}
		\max \{\epsilon_1^2 \lambda_1, \dots, \epsilon_N^2 \lambda_N \} \le \sum_{i=1}^{N} \epsilon_i^2 \lambda_i.
	\end{equation}
	
	Thus, we can find solutions in the complex field that have a cost function arbitrarily close to zero. However, since we need to enforce the integer constraint by rounding, these solutions will, for sufficiently small values of $\epsilon_i$, round to zero and not satisfy the full rank constraint $\mathbf{Z}$. We introduce the modified, relaxed version of (\ref{second_opt_form}) as
	
	\begin{equation}
	\label{alpha_opt_form}
		\begin{split}
			\mathbf{Z} = \underset{\mathbf{Z}_0}{\mathrm{argmin}} & \max_k \lvert \mathbf{z}_{0,k} \mathbf{G} \mathbf{z}_{0,k}^\textrm{H} - \frac{1}{\alpha^2}\rvert, \\
			& \textrm{s.t.} \ \textrm{rank}(\mathbf{Z}_0) = N.
		\end{split}
	\end{equation}
	
	 While (\ref{alpha_opt_form}) is still a nonconvex problem, we can however identify a set of $\mathbf{Z}_0$ matrices that reach the global minima of zero for the cost function as
	
	\begin{equation}
		\mathbf{Z}_0 = (\alpha \mathbf{Q}_0)^{-1} \mathbf{H},
	\end{equation}
	
	\noindent where $\mathbf{Q}_0$ is any unitary matrix. Given that $\mathbf{H}$ is full rank it follows immediately that $\mathbf{Z}_0$ is full rank as well.
	
	Once we have found a solution to the problem in (\ref{alpha_opt_form}) we now use a rounding procedure in order to ensure the integer constraint of $\mathbf{Z}$ and obtain its cost function. The entire procedure -- starting from the unitary matrix $\mathbf{Q}_0$ -- is termed \textit{Round}, consists of two different projections and is summarized in Table \ref{ap_algo}.
	
	We now discuss the projections involved in \textit{Round}. Projecting on the set of integer matrices of size $N \times N$ is achieved by rounding each element of the input matrix to the nearest integer. While rounding itself is straightforward, care must be taken in the case that we lose rank after the operation (e.g. an entire line or column is rounded to zero). 
	
	If this occurs, the projection on the set of full rank integer matrices of size $N \times N$ becomes non-trivial, possibly without a unique solution. For the purpose of our algorithm we define the projection operator as
	
	\begin{equation}
		P_{\textrm{Z}}(\mathbf{Z}_0) =
		 \begin{cases}
		 	\lceil \mathbf{Z}_0 \rfloor & \textrm{if rank} (\lceil \mathbf{Z}_0 \rfloor) = N, \\
		 	\mathbf{0}_{\textrm{N} \times  \textrm{N}} & \textrm{otherwise}.
		 \end{cases}
	\end{equation}

	The second projection ensures that the product constraint $\mathbf{H}_0 = \mathbf{Q}_0\mathbf{Z}_0$ is respected. Provided that both factors are full rank, this is achieved by inverting the integer matrix $\mathbf{Z}_0$ as
	
	\begin{equation}
		\mathbf{Q}_0 = P_{\mathbf{H}_0}(\mathbf{Z}_0) = \mathbf{H}_0 \mathbf{Z}_0^{-1}.
	\end{equation}
	
	Since LR aided MIMO detection requires the factorization to be exact and $\mathbf{Z}$ to be an integer matrix \cite{lr_determinant_discussion}, we can only relax the orthogonality condition on $\mathbf{Q}$, hence why the algorithm in Table \ref{ap_algo} terminates by projecting $\mathbf{Q}$ on the product constraint.
	
	\begin{table}
		\caption{The \textit{Round} Procedure}
		\label{ap_algo}
		\begin{algorithmic}[1]
			\renewcommand{\algorithmicrequire}{\textbf{Input:}}
			\renewcommand{\algorithmicensure}{\textbf{Output:}}
			
			\REQUIRE $\mathbf{H} \in \mathbb{C}^{\textrm{N} \times \textrm{N}}, \mathbf{Q}_0 \in \mathbb{C}^{\textrm{N} \times \textrm{N}}$
			\ENSURE  $\mathbf{Q} \in \mathbb{C}^{\textrm{N} \times \textrm{N}}, \mathbf{Z} \in (\mathbb{Z}+j\mathbb{Z})^{\textrm{N} \times \textrm{N}} \ \textrm{s.t.} \ \mathbf{H = QZ}$
			
			\textit{Initialize} $\mathbf{Z}$
			\STATE $\mathbf{Z} \leftarrow \mathbf{Q}^{-1} \mathbf{H}$
			
			\textit{Project on integer set}
			\STATE $\mathbf{Z} \leftarrow P_\textrm{Z}(\mathbf{Z})$

			\textit{Project on product constraint}
			\STATE $\mathbf{Q} \leftarrow P_{\mathbf{H}}(\mathbf{Z})$

			\RETURN $\mathbf{Q}$, $\mathbf{Z}$
		\end{algorithmic}
	\end{table}
	
	We call a solution produced by the \textit{Round} procedure feasible if the rounding projection does not degenerate. This is similar with the terminology used in \cite{boyd_noncvx}.
	
	From the entire collection of feasible solutions, we select the one with the lowest cost function as the final winner. Our simulation results will show that, given an adequate size of the search space, we can outperform the CLLL algorithm.
	
	We now describe the search strategy used to generate the collection of feasible solutions. We start by performing CLLL reduction on the channel matrix $\mathbf{H}$. This serves as a preprocessing step that helps increase the number of solutions found, since the CLLL is known to have an average effect of reducing the orthogonality defect of the input matrix \cite{lll_book}.
	
	Once the preprocessing is applied, we denote the resulted basis by $\mathbf{H}_0$ and let its singular value decomposition be
	
	\begin{equation}
		\mathbf{H}_0 = \mathbf{U} \mathbf{\Sigma} \mathbf{V}^\mathrm{H}.
	\end{equation}
	
	We define the matrix $\mathbf{Q}_1$ as the projection of $\mathbf{H}_0$ on the set of unitary matrices multiplied by a real, positive constant $\alpha$ i.e. $\mathbf{Q}_1 = \alpha \mathbf{U} \mathbf{V}^\textrm{H}$. Then, $\mathbf{Q}_1$ admits the unique eigen-decomposition
	
	\begin{equation}
		\mathbf{Q}_1 = \alpha \mathbf{V}_1 \mathbf{D} \mathbf{V}_1^\mathrm{H},
	\end{equation}
	
	\noindent where $\mathbf{D}$ is a diagonal matrix with the complex eigenvalues of $\mathbf{Q}_1$ as its diagonal elements. Since $\mathbf{Q}_1$ is a unitary matrix, the diagonal values of $\mathbf{D}$ are complex values that satisfy $\lvert d_{k,k}\rvert = 1$, with
	
	\begin{equation}
		\mathbf{D} =
		\begin{bmatrix}
		e^{-j \theta_1} & 0 & \dots  & 0 \\
		0 & e^{-j \theta_2} & \dots  & 0 \\
		\vdots & \vdots & \ddots & \vdots \\
		0 & 0 & \dots  & e^{-j \theta_N}
		\end{bmatrix}.
	\end{equation}
	
	For the purpose of our algorithm, we fix the $\mathbf{V}_1$ matrix and parametrize $\mathbf{Q}_1$ by the angles of its eigenvalues and the amplitude factor $\alpha$ as the function
	
	\begin{equation}
		\mathbf{Q}_1 = f(\alpha, \theta_1, \theta_2, \dots, \theta_N).
	\end{equation}
	
	The corresponding solution $\mathbf{Z}_0$ is obtained by recognizing that any $\mathbf{Q}_1$ parametrized in such a way is a solution to the relaxed problem (\ref{alpha_opt_form}), thus we can apply the \textit{Round} procedure as
	
	\begin{equation}
		\mathbf{Z}_0 = \textit{Round}(f(\alpha, \theta_1, \dots, \theta_N), \mathbf{H}_0).
	\end{equation}
	
	The optimization problem that must be solved is now
	
	\begin{equation}
		\label{full_exponential_sweep}
		\mathbf{Z} = \underset{\alpha, \theta_1, \theta_2, \dots, \theta_N}{\mathrm{argmin}} c(\mathbf{Z}_0 \mathbf{H}_0^{-1}).
	\end{equation}
	
	Instead of searching across the entire set of unitary matrices of size $N$ -- with a number of $N^2-1$ degrees of freedom \cite{golub_alg} -- for feasible solutions, we thus restrict our candidate set to a number of $N$ dimensions plus one dimension for the $\alpha$ constant.
		
	The problem in (\ref{full_exponential_sweep}) is still highly nonconvex and the only guaranteed way of finding its global minimum is by jointly searching across all $N+1$ degrees of freedom with a sufficiently small granularity.
	
	We now further reduce the complexity of the problem in (\ref{full_exponential_sweep}) by jointly searching across pairs of dimensions $(\alpha, \theta_i)$ for all $i \in \{ 1, \dots, N \}$, instead of the entire space $(\alpha, \theta_1, \dots, \theta_N)$. This loses the guarantee of a globally optimum solution, but our numerical simulations indicate the loss is negligible and that this is a suitable heuristic. The optimization problem becomes
	
	\begin{equation}
		\label{final_optimization_eq}
		\mathbf{Z} = \ \underset{\alpha, \theta_i}{\mathrm{argmin}} \ c(\mathbf{Z}_0 \mathbf{H}^{-1}), \ \text{for} \ i \in \{1, \dots, N\}.
	\end{equation}
	
	In order to solve (\ref{final_optimization_eq}) we resort to performing $N$ concurrent exhaustive searches in the $(\alpha, \theta_i)$ spaces. Furthermore, since $\alpha$ and $\theta_i$ are continuous variables, we uniformly sample the sets $A$ and $\Theta$ for their values.
	
	Once the search is completed across all two-dimensional spaces $(\alpha, \theta_i)$ the $\mathbf{Z}$ matrix with the lowest cost function is selected as winner and the corresponding $\mathbf{Q}$ is derived as $\mathbf{Q} = \mathbf{H} \mathbf{Z}^{-1}$.
	
	In case two or more solutions exhibit the same minimum cost function, we use the sum-noise amplification $s(\mathbf{Z})$ as a metric to decide between them. This has the expression
	
	\begin{equation}
		s(\mathbf{Z}) = \sum_{k} \mathbf{z}_k \mathbf{G} \mathbf{z}
	\end{equation}
	
	\noindent and is justified from a MIMO detection perspective. Alternatively, one could decide based on the second-highest noise amplification metric, but our simulations suggested that the difference in performance is negligible.
	
	The proposed algorithm is summarized in Table \ref{main_algo}.
	
	Finally, we note that a very similar method of relaxing the optimization problem, identifying feasible solutions and rounding can be applied when the cost function is the orthogonality defect in (\ref{od_metric}). However, it turns out that the respective metric is invariant up to a scalar factor $\alpha$, making it unclear how to select between two solutions with the same metric.
	
	\begin{table}
		\caption{The Proposed Lattice Basis Reduction Algorithm}
		\label{main_algo}
		\begin{algorithmic}[1]
			\renewcommand{\algorithmicrequire}{\textbf{Input:}}
			\renewcommand{\algorithmicensure}{\textbf{Output:}}
			\REQUIRE $\mathbf{H} \in \mathbb{C}^{\textrm{N} \times \textrm{N}}, N_{\alpha}, N_{\theta}, A = [1, \frac{N}{2}], \Theta = [0, 2\pi]$
			\ENSURE  $\mathbf{Q} \in \mathbb{C}^{\textrm{N} \times \textrm{N}}, \mathbf{Z} \in (\mathbb{Z}+j\mathbb{Z})^{\textrm{N} \times \textrm{N}} \ \textrm{s.t.} \ \mathbf{H = QZ}$
			\\ \textit{Preprocessing} :	
			\\ $\mathbf{H}_0 \leftarrow \textrm{CLLL}(\mathbf{H})$
			\\ $[\mathbf{U}, \mathbf{S}, \mathbf{V}] = \textrm{svd}(\mathbf{H}_0)$
			\\ $\mathbf{Q}_1 \leftarrow \mathbf{U} \mathbf{V}^\textrm{H}$
			\\ $[\mathbf{V}_1, \mathbf{D}] = \textrm{eig}(\mathbf{Q}_0)$
			
			\FOR {$i = 1$ to $\textrm{N}$}
			\FOR {$j = 1$ to $\textrm{N}_{\alpha}$}
			\STATE $\alpha \leftarrow \textrm{A}[j]$
			\FOR {$k = 1$ to $\textrm{N}_{\theta}$}
			\STATE $d_{i,i} \leftarrow e^{-j\Theta[k]}$
			\STATE $\mathbf{Q}_1 \leftarrow \alpha \mathbf{V}_1 \mathbf{D} \mathbf{V}_1^{\textrm{H}}$
			\\ \textit{Round procedure}
			\STATE $\mathbf{Z} \leftarrow \textit{Round}(\mathbf{Q}_1, \mathbf{H}_0)$
			\\ \textit{Evaluate cost function}
			\STATE $\textrm{cost}(i,j,k) = c(\mathbf{Z}\mathbf{H}_0^{-1})$
			\ENDFOR
			\ENDFOR
			\ENDFOR
			\\ \textit{Select solution with lowest cost function}
			\STATE $[i_0, j_0, k_0] = \mathrm{argmin} (\textrm{cost}(i,j,k))$
			\\ \textit{Reconstruct winning solution}
			\STATE $\mathbf{Z} \leftarrow \textit{Round}(\mathbf{Q}_1(i_0, j_0, k_0), \mathbf{H}_0)$
			\STATE $\mathbf{Q} \leftarrow \mathbf{H}_0 \mathbf{Z}^{-1} $
			\STATE $\mathbf{Z} \leftarrow \mathbf{Q}^{-1} \mathbf{H}$
			\RETURN $\mathbf{Q}$, $\mathbf{Z}$
		\end{algorithmic}
	\end{table}
	
	\section{Performance Results and Complexity}
	\label{section_perf}
	\subsection{Parameter Choice}
	We describe the choice of range and granularity when searching in the $(\alpha, \theta_i)$ spaces.
	
	Since $e^{-j \theta_i}$ is periodic with period $2\pi$, we can restrict our search for $\theta_i$ by uniformly sampling  $N_\theta$ points in the $[0, 2\pi)$ interval. Our experiments show that $N_\theta = 30$ offers a sufficient granularity to cover a large number of feasible solutions, regardless of the matrix dimension $N$.
	
	We now explain the effect of the $\alpha$ value. Since projecting on the set of integers is a nonlinear operator, we can exploit this in order to find new factorizations of the original matrix by observing that using $\alpha \mathbf{Q}_1$ as a solution to the relaxed problem (\ref{alpha_opt_form}) is the same as factorizing the scaled matrix $\frac{1}{\alpha} \mathbf{H}$ using $\mathbf{Q}_1$ as initial point. This leads us to two solutions:
	
	\begin{equation}
	\begin{gathered}
	\mathbf{H = QZ}, \\
	\frac{1}{\alpha} \mathbf{H} = \mathbf{Q}_a \mathbf{Z}_a \implies \mathbf{H} = \alpha \mathbf{Q}_a \mathbf{Z}_a.
	\end{gathered}
	\end{equation}

	If $\alpha$ is sufficiently smaller or larger than 1, the two solutions $\mathbf{Z}$ and $\mathbf{Z}_a$ will be different because of the nonlinear behaviour of the rounding function. Additionaly, the cost function of the scaled solution will be proportional to $\frac{1}{\alpha^2}$, since	
	
	\begin{equation}
		(\alpha \mathbf{Q}_a)^{-1} (\alpha \mathbf{Q}_a)^{-\mathrm{H}} = \frac{1}{\alpha^2} \mathbf{Q}_a^{-1} \mathbf{Q}_a^{-\mathrm{H}}.
	\end{equation}
	
	Thus, choosing $\alpha > 1$ can lead to much better factorizations. For each eigenvalue $\theta_i$ we search for the optimal $\alpha$ value by uniformly sampling $N_\alpha$ points in the $[1, N/2]$ interval for an input matrix of size $N$. Our experiments show that, at least for $N \le 8$, $N_\alpha = 20$ is an acceptable granularity.

	\subsection{Performance Results}
	Once the parameters are set, the algorithm is applied to matrices with i.i.d. elements drawn from a standard complex normal distribution. We compare the performance of our algorithm with that of the CLLL algorithm in \cite{clll} by the ratio of the two cost functions
	
	\begin{equation}
		\textrm{R} = \frac{c(\mathbf{Z}_\textrm{clll} \mathbf{H}^{-1})}{c(\mathbf{Z}_\textrm{prop} \mathbf{H}^{-1})}
		= \frac{ \max\limits_k \ (\mathbf{Z}_{\textrm{clll}} \mathbf{H}^{-1} \mathbf{H}^{-\textrm{H}} \mathbf{Z}_{\textrm{clll}}^{\textrm{H}})_{k,k}} {\max\limits_k \ (\mathbf{Z}_{\textrm{prop}} \mathbf{H}^{-1} \mathbf{H}^{-\textrm{H}} \mathbf{Z}_{\textrm{prop}}^{\textrm{H}})_{k,k}},
	\end{equation}
	
	\noindent where $\mathbf{Z}_\textrm{clll}$ and $\mathbf{Z}_\textrm{prop}$ are the outputs of the CLLL and the proposed algorithm, respectively.
	
\begin{figure}[!t]
		\centering
		\includegraphics[width=2.5in]{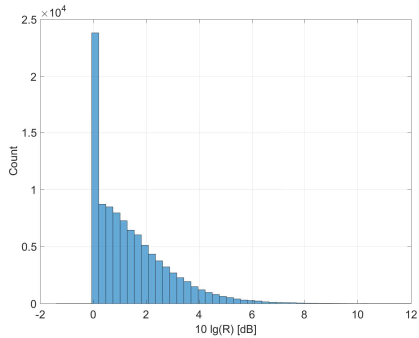}
		\caption{The empirical distribution of $10\lg\text{R}$ after running the proposed algorithm on 100,000 matrices of size $N = 4$.}
		\label{ratioN4}
	\end{figure}

\begin{figure}[!t]
		\centering
		\includegraphics[width=2.5in]{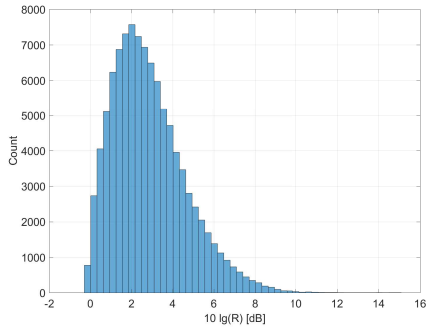}
		\caption{The empirical distribution of $10\lg\text{R}$ after running the proposed algorithm on 100,000 matrices of size $N = 8$.}
		\label{ratioN8}
	\end{figure}

	\figurename{} \ref{ratioN4} and \ref{ratioN8} plot the empirical distribution of $10\lg \textrm{R}$ for $N = \{4, 8\}$ respectively. There is a strong mode at 0 dB in the distribution for $N = 4$ corresponding to the cases where the algorithm produces the same solution as CLLL -- roughly 17\%. However, for $N=8$ this mode quickly vanishes. Furthermore, observe that for $N=4$ the algorithm never produces worse results than the initial, CLLL preprocessed matrix.
	
	Based on \figurename{} \ref{ratioN8}, we conclude that the output of the proposed algorithm is significantly different from that of the CLLL for increasing $N$. This is also reflected in Table \ref{det_table}, where we show the empirical distribution of $\lvert \det \mathbf{Z}_\textrm{prop} \rvert$. Since the only constraint we place on $\mathbf{Z}_\textrm{prop}$ is that its entries are Gaussian integers, we do not satisfy the unimodularity constraint for some solutions.
	
	\begin{table}[!t]
		\renewcommand{\arraystretch}{1.4}
		\caption{Determinant and Coverage of the Proposed Algorithm for a Total of $100,000$ Matrices}
		\label{det_table}
		\centering
		\begin{tabular}{|c|c|c|c|c|c|c|c|}
			\hline
			\multirow{2}{*}{} & \multicolumn{6}{|c|}{$\lvert \det \mathbf{Z}_\textrm{prop} \rvert \ [\%]$} & %
			\multirow{2}{*}{\textbf{Coverage}}\\
			\cline{2-7}
			\cline{2-7}
			& $1$ & $\sqrt{2}$ & $2$ & $\sqrt{5}$ & $2\sqrt{2}$ & $> 2\sqrt{2}$ & [\%] \\
			\hline
			$\boldsymbol{N = 4}$ & $90.6$ & $5.4$ & $2$ & $0.5$ & $0.2$ & $1.3$ & $99.28$\\
			\hline
			$\boldsymbol{N = 8}$ & $50$ & $12.7$ & $10.1$ & $4.3$ & $3.1$ & $19.8$ & $99.99$ \\
			\hline
		\end{tabular}
	\end{table}
	
	Although it may seem that not satisfying this constraint leads to a performance loss in MIMO detection, our simulations show the opposite effect occurs: the performance gap increases with $N$, while the distribution of $\lvert \det \mathbf{Z}_\textrm{prop} \rvert$ shifts away from unit value. This observation is also referenced in \cite{lr_determinant_discussion}, where the authors mention that the unimodularity of $\mathbf{Z}$ is not in fact a required condition.
	
	Table \ref{det_table} also shows the coverage of the proposed algorithm, representing the percentage of matrices for which at least one feasible solution is found during the greedy search.
	
	Once we have generated a sufficient number of factorizations, we can investigate the performance gain in a LR-aided MIMO detection scenario. We note that while the cost function we have optimized over is a good indicator of the performance of a MIMO detection algorithm \cite{lr_determinant_discussion}, it can only be used as an upper bound.
	
	We use the proposed basis reduction algorithm in a MIMO detector that employs a LR-aided ZF strategy and compare it with the version that uses the CLLL as a baseline. For each channel matrix $\mathbf{H}$ we simulate a number of $100$ transmissions and plot the uncoded BER, where the transmitted signal belongs to a \{4, 16\}-QAM constellation.
	
\begin{figure}[!t]
		\centering
		\includegraphics[width=2.5in]{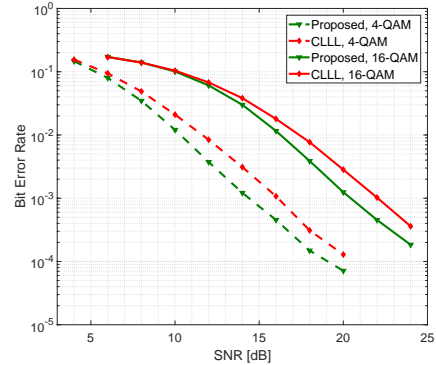}
		\caption{The uncoded BER performance of a MIMO receiver using the proposed algorithm for lattice basis reduction averaged across 100,000 channels for $N = 4$.}
		\label{berN4}
	\end{figure}

\begin{figure}[!t]
		\centering
		\includegraphics[width=2.5in]{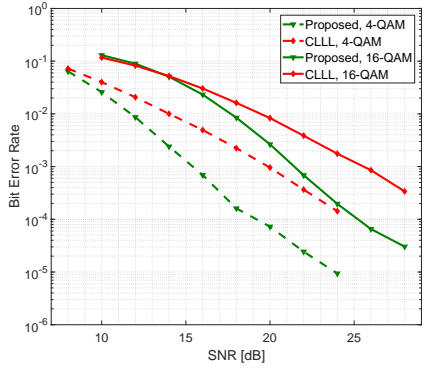}
		\caption{The uncoded BER performance of a MIMO receiver using the proposed algorithm for lattice basis reduction averaged across 100,000 channels for $N = 8$.}
		\label{berN8}
	\end{figure}

	\figurename{} \ref{berN4} and \ref{berN8} show the BER results for scenarios with $N = 4$ and $N = 8$ respectively. We conclude that our algorithm outperforms the CLLL by at least 1.5 dB in the high SNR regime for the $N = 4$ scenario and by at least 3 dB for $N = 8$. Furthermore, we notice that the performance gap for the $N = 8$ case increases with SNR, suggesting a higher diversity order is achieved.
	
\begin{figure}[!t]
		\centering
		\includegraphics[width=2.5in]{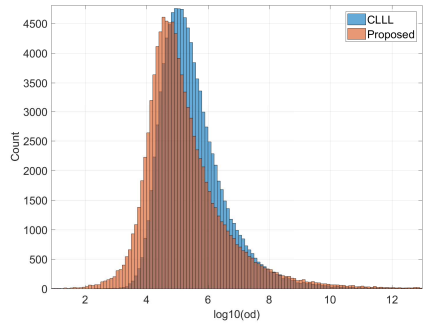}
		\caption{The empirical distribution of lg(\textit{od}), where \textit{od} is the orthogonality defect measure in (\ref{od_metric}) for CLLL and the proposed algorithm when $N=8$.}
		\label{od_N8_fig}
	\end{figure}

	We also investigate the performance of our algorithm in terms of the orthogonality defect metric introduced in (\ref{od_metric}). \figurename{} \ref{od_N8_fig} shows the distribution of the orthogonality defect metric of the outputs produced by the proposed algorithm and the CLLL, respectively for $N=8$, where we notice that a decrease in this metric is achieved as well.
	
	\subsection{Complexity Analysis}
	The complexity of the innermost loop of the algorithm is given by the floating point operation (FLOP) count of the \textit{Round} procedure. Since our solutions $\mathbf{Q}_0$ to the relaxed problem are unitary matrices, inverting them is trivial and amounts to a matrix multiplication. While the rounding part in the integer projection is trivial, the rank check requires performing an SVD decomposition. The final step consists in inverting an integer matrix, for which we assume no special structure and treat as a regular matrix inversion.
	
	The complexity of the proposed algorithm measured in complex FLOPs is summarized in Table \ref{complx_table}.	Note that the CLLL preprocessing step is not included in the analysis because of its non-deterministic cost. Average FLOP counts for the CLLL algorithm can be found in \cite[Table II]{clll}, however its asymptotic cost is known to be of the order $\mathcal{O}(N^4 \log B)$, with $B$ the norm of the longest column of the input matrix.
		
	\begin{table}[!t]
		\renewcommand{\arraystretch}{1.4}
		\caption{FLOP Count of the Proposed Algorithm}
		\label{complx_table}
		\centering
		\begin{tabular}{|c|c|}
			\hline
			\bfseries Operation & \bfseries FLOP Count \\
			\hline
			$P_\textrm{Z}$ (Rounding $\mathbf{Z}$ and rank check) & $\frac{8}{3} N^3$ \\
			\hline
			$P_\mathbf{H}$ (Inverting $\mathbf{Q}_0$ and $\mathbf{Z}$) & $N^3 + \frac{4}{3} N^3$\\
			\hline
			$Round$ & $5 N^3$ \\
			\hline
			\hline
			\textbf{CLLL} \cite{clll} & $\mathcal{O}(N^4 \log B)$ \\
			\hline
			\textbf{Proposed algorithm} & $5 N_\theta N_\alpha N^4 \approx \mathcal{O}(N^4)$ \\
			\hline
		\end{tabular}
	\end{table}
	
	The proposed algorithm is thus of the same polynomial order as the CLLL. Besides the FLOP count, our algorithm has the advantage that it is inherently suited for parallelization by distributing the search across multiple $(\alpha, \theta_i)$ spaces.
	
	Ignoring the $\alpha$ factor, all the solutions for the relaxed problem (\ref{final_optimization_eq}) can be expressed in the form $\mathbf{Q}_0 = \mathbf{V} \mathbf{D} \mathbf{V}^\textrm{H}$, forming a closed group under matrix multiplication. Thus, the entire set of matrices that sweep a given eigenvalue can be obtained from the original $\mathbf{Q}_0$ by multiplication with matrices of the form $\mathbf{V} \mathbf{D}_{i} \mathbf{V}^\textrm{H}$, where $\mathbf{D}_{i}$ is the identity matrix with the $i$-th diagonal element replaced by $e^{-j\theta_i}$.
	
	\section{Conclusions}
	\label{section_conc}
	
	In this paper we have proposed a novel algorithm for complex lattice basis reduction and demonstrated its performance in a lattice-reduction aided MIMO detection scenario. By taking an optimization perspective, we have introduced a relaxed version of the lattice basis reduction problem for which a family of global solutions is easily identified.
	
	By further applying a greedy search across the amplitude and eigenvalue dimensions, we are able to select a winning solution. Our simulations show that we can outperform the CLLL algorithm with a performance gap increasing with the matrix size $N$. The complexity analysis has shown that the FLOP count of the proposed algorithm is of the same polynomial order as CLLL reduction and is highly suited for a parallelized implementation.
	
	Future research directions include developing theoretical justifications and guarantees of the search heuristic we have used, as well as extending the proposed framework to other nonconvex optimization problems.
	
	\bibliographystyle{IEEEtran}
	\bibliography{myBib}
	
\end{document}